# Evaluation of a permeability-porosity relationship in a low permeability creeping material using a single transient test

Siavash Ghabezloo[*(1)], Jean Sulem[(1)], Jérémie Saint-Marc[(2)]

(1) *Université Paris-Est, UR Navier, CERMES, Ecole Nationale des Ponts et Chaussées, Marne la Vallée, France*
(2) *TOTAL, Management of Residual Gases Project, Pau, France*

## Abstract

A method is presented for the evaluation of the permeability-porosity relationship in a low-permeability porous material using the results of a single transient test. This method accounts for both elastic and non-elastic deformations of the sample during the test and is applied to a hardened class G oil well cement paste. An initial hydrostatic undrained loading is applied to the sample. The generated excess pore pressure is then released at one end of the sample while monitoring the pore pressure at the other end and the radial strain in the middle of the sample during the dissipation of the pore pressure. These measurements are back analysed to evaluate the permeability and its evolution with porosity change. The effect of creep of the sample during the test on the measured pore pressure and volume change is taken into account in the analysis. This approach permits to calibrate a power law permeability-porosity relationship for the tested hardened cement paste. The porosity sensitivity exponent of the power-law is evaluated equal to 11 and is shown to be mostly independent of the stress level and of the creep strains.

[*]Corresponding Author: Siavash Ghabezloo, CERMES, Ecole Nationale des Ponts et Chaussées, 6-8 avenue Blaise Pascal, Cité Descartes, 77455 Champs-sur-Marne, Marne la Vallée cedex 2, France,
Email: ghabezlo@cermes.enpc.fr

# 1  Introduction

The evaluation of the permeability in the laboratory is basically an inverse problem. The two most widely used methods for evaluation of the permeability of geomaterials are the steady state method and the transient pulse method. The steady state method consists in applying a constant pressure gradient to the sample and measuring the resulted flow rate. The permeability of the sample is then calculated using Darcy's law. This method is particularly appropriate for high permeability materials. The transient pulse method is based on the evaluation of the decay of a small step change of pressure imposed in one end of a sample. This method, which was pioneered by Brace *et al*. [1], is appropriate for low permeability materials and is used extensively to evaluate the permeability of different geomaterials (e.g. Bernabé [2][3] on granite, Escoffier *et al*. [4] on mudstone). In the original setting of Brace *et al*. [1], a cylindrical sample is connected to two fluid reservoirs with equal pressures. A sudden increase of the fluid pressure in the upstream reservoir unbalances the system and causes a fluid flow from this reservoir, through the sample to the downstream reservoir in order to equilibrate the pressures in the sample and in the reservoirs. The permeability of the sample is then back analysed from the kinetics of the decay of the pressure in the upstream reservoir. Some alternative methods to evaluate the permeability of low permeability materials have been recently developed by monitoring the response of a saturated body to mechanical and thermal strains. Among them, the beam bending method [5][6][7] is based on the fact that bending of a saturated porous beam creates a pressure gradient in the pores, as the top half is in compression and the half bottom is in tension. This pressure gradient causes the fluid flow to equilibrate the pressure and consequently the force required to sustain a fixed deflection decreases. The permeability and also the elastic modulus of the body are evaluated by analysing the kinetics of force relaxation. Thermopermeametry is another method that is based on the analysis of thermal expansion kinetics [6][7][8]. The thermal expansion of water is higher than the one of solids, so a rapid temperature increase and then isothermal hold of a saturated porous body causes an initial dilation and pore fluid pressurization due to the greater expansion of the pore fluid. This initial dilation is then followed by a contraction of the body caused by the dissipation of the excess pore pressure and the fluid outflow. Analysis of the kinetics of thermal dilation yields the permeability of the body. Beam bending and thermopermeametry methods were originally developed for gels and later extended to more rigid materials like cement paste and mortar. Dynamic pressurization [9][10] is another permeability evaluation method which is performed on a sample that is enclosed in a vessel full of fluid under a constant pressure. In the initial state, the pressure in the vessel and the sample pore pressure are in equilibrium. A sudden increase of the pressure in the vessel causes an initial contraction of the sample. This contraction, when the pressure is kept constant, is followed by a time dependent dilation due to the progressive increase of the sample pore pressure to reach equilibrium with the pressure inside the vessel. The permeability is then evaluated by analysing the kinetics of the dilation of the sample.

One can see that these methods used for evaluation of transport properties of low permeability materials are based on the analysis of the kinetics of diffusion of pore pressure or the strains induced by pore pressure diffusion. The same principle is used in this work for the evaluation of the permeability of a hardened cement paste. The test method used here is similar to the one presented by Hart and Wang [11] as a single test method for evaluation of poroelastic constants and flow parameters of low permeability rocks. These authors used the excess pore pressure generated in a sample due to an undrained loading for the evaluation of the permeability. The pressure at one side of the sample is connected to a reservoir with a constant volume and the kinetics of the pore pressure variations at the other end of the sample and in the reservoir is analysed for evaluation of the permeability.

Our experimental study for the evaluation of the permeability of a hardened cement paste is a part of a larger study on the thermo-poro-mechanical behaviour of this material for petroleum applications. Indeed, in oil wells, a cement sheath is placed between the rock and the casing for support and sealing purpose. The cement lining is submitted to various thermal and mechanical loadings during the life of the well from the drilling phase to the production phase and finally in the abandonment phase when the well must seal the subsurface from the surface, as for instance for storage and sequestration of greenhouse gas. In due course of these actions, the cement can be damaged and the mechanical and transport properties can be degraded, this degradation being detrimental to its functions. Moreover, the determination of the permeability of this cement over time is essential for the prediction of the sealing performance of the well when $CO_2$ storage and sequestration is planned. The results of our previous experimental study including drained and undrained hydrostatic compression tests, unjacketed tests and drained and undrained heating tests have been presented in Ghabezloo *et al*. [12][13].

By applying an undrained hydrostatic loading, an excess pore pressure is generated inside the sample, related to the applied hydrostatic stress by the Skempton coefficient of the material. At the end of the hydrostatic loading phase, a constant pore pressure much lower than the existing pore pressure inside the sample is applied at one end of the sample. Under this pressure boundary condition, the pore fluid flows out of the sample and the pore pressure decreases. The pore pressure change at the other end of the sample and the radial strain in the middle of the sample are both measured as functions of time and back analysed to evaluate the permeability. During the test, the pore pressure in the sample varies greatly from one end to another. This variable pore pressure keeping the confining pressure constant, induces a variable effective stress in the sample and results in a heterogeneous strain field. The stress-dependent character of the poroelastic parameters of the hardened cement paste (Ghabezloo *et al.* [12]) and also the creep of the material during the test add some particular aspects to the back-analysis, which makes this problem different from the classical solutions of transient permeability evaluation tests. The porosity changes due to the increase of the effective stress and also the presence of additional creep deformations induce a decrease of the permeability during the test. The coupled measurement of pore pressure and deformations during the test is back analysed to evaluate the permeability-porosity relationship using the results of a single transient test.

It is well-known that there is no unique permeability-porosity relationship that can be applied to all porous materials. As mentioned by Bernabé *et al*. [14], one reason is that the porosity is a scale invariant material property; if the material and its pore space could be uniformly expanded or contracted everywhere, there would be no change in the porosity, but the permeability would change in this case. Moreover, different pores in a given material have different contributions to the permeability of the material according to their size and shape. The porosity takes into account only the relative volume of the pores to the total volume, and not the shape and the distribution of the size of the pores. Thus two porous materials with the same porosity, but different pore shape and pore-size distributions, must have different permeabilities. For a given evolution process that changes both permeability and porosity of a porous material, for example elastic or plastic compaction, microcracking or chemical alteration, it is usually assumed that there is a power-law relationship $k \propto \phi^\alpha$ between these parameters (Bernabé *et al*. [14]). The exponent of this relation may be integer or non-integer, constant or variable, according to the properties of the material and of the evolution process. Based on the experimental data of Bernabé *et al* [15], Walder and Nur [16] postulated a slightly different power-law in which the permeability vanishes at a critical non-zero porosity. David *et al*. [17] presented a compilation of published data on the permeability-porosity relationships for different geomaterials. One can see that the reported values of exponent $\alpha$ vary between 1.1 and 25.4 for different materials and the higher values correspond in general to rocks with a high porosity. However, no clear correlation between $\alpha$ and the petrophysical properties could be found. Experimental study of Zhu and Wong [18] on Berea and Boise sandstones resulted in a variable $\alpha$ which increases when pressure increases. Meziani [19] performed an experimental study on the gas permeability of mortar under hydrostatic loading and found a strongly non-linear permeability-porosity relationship. The exponent $\alpha$ is found to be a decreasing function of the loading level, decreasing from 36 to 26 when the confining pressure is increased up to 57 MPa. Based on an idea presented by Bernabé *et al*. [14], the total porosity of the rock can be separated into two categories of effective and non-effective porosity according to the contribution of the pores in the fluid transport. These two quantities are not purely geometrical and depend on the fluid velocity field. These authors propose that the exponent $\alpha$ is related to the ratio of the effective to non-effective porosity and its variation during an evolution process so that $\alpha$ is dependent upon the particular physical process of porosity evolution and the pore geometry of the material. Consequently as emphasized by Bernabé *et al*. [14], searching a permeability-porosity relationship is meaningful either for a given rock sample during a porosity evolution process, for example elastic compaction, or for different rock samples only if these samples correspond to different stages of the same evolution process.

## 2 Poromechanical background

We present here the framework used to describe the response of the sample during the performed test. The equations are written for a porous material which is not necessarily homogeneous and isotropic at the micro-scale. The theoretical basis of the formulation has been presented in many earlier studies. Among them, one can refer to the milestone papers and textbooks of Biot and Willis [22], Brown and Korringa [23], Rice and Cleary [24], Zimmerman [25], Detournay and Cheng [26], Vardoulakis and Sulem [27], Wang [28], Coussy [29]. This framework is recalled here in a comprehensive manner in order to clarify the mathematical and physical significance of the different parameters used in the analysis and also to take into account the effect of non-elastic strains in the evaluation of the permeability.

### *2.1 Poroelastic formulation*

The porosity $\phi$ of a porous material is defined as the ratio of the volume of the porous space $V_\phi$ to total volume $V$ in the actual (deformed) state:

$$\phi = \frac{V_\phi}{V} \quad (1)$$

We consider a saturated sample under an isotropic state of stress $\sigma$ (positive in compression) and we define the differential pressure $\sigma_d$ (i.e. Terzaghi effective stress) as the difference between the confining pressure $\sigma$ and the pore pressure $p_f$.

$$\sigma_d = \sigma - p_f \quad (2)$$

The variations of the total volume $V$ and of the pore volume $V_\phi$ are given in the following expressions as a function of variations of two independent variables, $\sigma_d$ and $p_f$:

$$\frac{dV}{V} = -\frac{d\sigma_d}{K_d} - \frac{dp_f}{K_s}$$

$$\frac{dV_\phi}{V_\phi} = -\frac{d\sigma_d}{K_p} - \frac{dp_f}{K_\phi}$$

(3)

Where $K_d$, $K_s$, $K_p$ and $K_\phi$ are four elastic moduli defined below:

$$\frac{1}{K_d} = -\frac{1}{V}\left(\frac{\partial V}{\partial \sigma_d}\right)_{p_f} \quad , \quad \frac{1}{K_p} = -\frac{1}{V_\phi}\left(\frac{\partial V_\phi}{\partial \sigma_d}\right)_{p_f}$$

(4)

$$\frac{1}{K_s} = -\frac{1}{V}\left(\frac{\partial V}{\partial p_f}\right)_{\sigma_d} \quad , \quad \frac{1}{K_\phi} = -\frac{1}{V_\phi}\left(\frac{\partial V_\phi}{\partial p_f}\right)_{\sigma_d}$$

(5)

Equation (4) corresponds to an isotropic drained compression test in which the pore pressure is kept constant inside the sample. The variations of the total volume of the sample $V$ and of the volume of the pore space $V_\phi$ with the applied confining pressure give respectively the drained bulk modulus $K_d$ and the modulus $K_p$. Equation (5) corresponds to the unjacketed compression test, in which equal increments of confining pressure and pore pressure are simultaneously applied to the sample. The differential pressure $\sigma_d$ in this condition remains constant. The variation of the volume of the sample with respect to the applied pressure gives the unjacketed modulus $K_s$. The variation of the pore volume of the sample in this test could in principle be used to evaluate the modulus $K_\phi$. However experimental evaluation of this parameter is very difficult as the volume of the fluid exchanged between the sample and the pore pressure generator has to be corrected for the effect of fluid compressibility, and also for the effect of the deformations of the pore pressure generator and of the drainage system in order to access to the true variation of the pore volume of the sample. In the case of a porous material which is homogeneous and isotropic at the micro-scale $K_s = K_\phi = K_m$, where $K_m$ is the bulk modulus of the single solid constituent of the porous material. In the case of a porous material which is composed of two or more solids and therefore is heterogeneous at the micro-scale, the unjacketed modulus $K_s$ is some weighted average of the bulk moduli of solid constituents [30]. The modulus $K_\phi$ for such a material has a complicated dependence on the material properties. Generally it is not bounded by the elastic moduli of the solid components and can even have a negative sign if the bulk moduli of the individual solid components are greatly different one from another [31][32]. From Betti's reciprocal theorem, the following relation holds between the elastic moduli [23][33]:

$$\frac{1}{K_p} = \frac{1}{\phi}\left(\frac{1}{K_d} - \frac{1}{K_s}\right)$$

(6)

Using equation (6), the number of the required parameters to characterize the volumetric poro-elastic behaviour of a porous material is reduced to three, and among them, the experimental evaluation of the modulus $K_\phi$ is very difficult. Using the definition of the porosity presented in equation (1), the following equation is obtained for the variation of the porosity:

$$\frac{d\phi}{\phi} = \frac{dV_\phi}{V_\phi} - \frac{dV}{V}$$

(7)

Replacing equation (3) and then equation (6) in equation (7), the expression of the variation of porosity is found:

$$\frac{d\phi}{\phi} = -\frac{1}{\phi}\left(\frac{1-\phi}{K_d} - \frac{1}{K_s}\right)d\sigma_d + \left(\frac{1}{K_s} - \frac{1}{K_\phi}\right)dp_f \tag{8}$$

As shown by Hart and Wang [34], for the case of the axisymmetric triaxial test, the transient pore pressure can be approximated by the solution obtained for the one-dimensional diffusive flow except at early times for which the fully coupled poroelastic response includes also radial flow. Thus, for the analysis of the test results, the one-dimensional solution is used in the following.

Consider now an elementary volume with the $z$ axis parallel to flow direction, $q$ the fluid mass flux per surface area $S$ (orthogonal to $z$) and $M_f$ the total fluid mass inside the volume. Fluid mass conservation implies that:

$$\frac{1}{V}\frac{dM_f}{dt} + \frac{\partial q}{\partial z} = 0 \tag{9}$$

Let's define $m_f$ as the fluid mass per unit volume of the porous material, $m_f = M_f/V = \rho_f \phi$. The variations of $m_f$ is written as:

$$\frac{dm_f}{dt} = \frac{1}{V}\frac{dM_f}{dt} - \frac{M_f}{V}\frac{dV/V}{dt} \tag{10}$$

Replacing equation (10) in equation (9) we obtain:

$$\frac{dm_f}{dt} + m_f\frac{dV/V}{dt} + \frac{\partial q}{\partial z} = 0 \tag{11}$$

Knowing that $m_f = \rho_f \phi$ we can write $dm_f = \phi d\rho_f + \rho_f d\phi$. Replacing $d\rho_f = \rho_f dp_f / K_f$ and equation (7) in this relation we obtain:

$$dm_f = \phi \rho_f \frac{dp_f}{K_f} + \phi \rho_f \left(\frac{dV_\phi}{V_\phi} - \frac{dV}{V}\right) \tag{12}$$

Inserting equation (12) and $m_f = \rho_f \phi$ in equation (11) the following expression is obtained:

$$\phi \rho_f \left(\frac{1}{K_f}\frac{dp_f}{dt} + \frac{dV_\phi/V_\phi}{dt}\right) + \frac{\partial q}{\partial z} = 0 \tag{13}$$

Replacing equation (3) and then equation (6) in equation (13) the following relation is found for the case of $d\sigma = 0$:

$$\rho_f\left[\phi\left(\frac{1}{K_f} - \frac{1}{K_\phi}\right) + \left(\frac{1}{K_d} - \frac{1}{K_s}\right)\right]\frac{dp_f}{dt} + \frac{\partial q}{\partial z} = 0 \tag{14}$$

The fluid mass flux $q$ in equation (14) is given by Darcy's law:

$$q = -k\frac{\rho_f}{\mu_f}\frac{\partial p_f}{\partial z} \tag{15}$$

where $k$ is the permeability and $\mu_f$ is the fluid viscosity. Replacing equation (15) in equation (14) the following expression is obtained:

$$\frac{dp_f}{dt} = \frac{\beta_u}{\rho_f} \frac{\partial}{\partial z}\left(k \frac{\rho_f}{\mu_f} \frac{\partial p_f}{\partial z}\right) \tag{16}$$

where:

$$\beta_u = \frac{1}{\phi(1/K_f - 1/K_\phi) + (1/K_d - 1/K_s)} \tag{17}$$

## 2.2 Effect of non-elastic strains

The above framework can be extended to account for the effect of non-elastic strains which can be produced during the test. These strains can be plastic, viscoelastic or viscoplastic and induce non-elastic porosity changes. The non-elastic changes of the total volume, pore volume and solid volume are defined by:

$$dV^{ne} = dV - dV^e \,;\, dV_\phi^{ne} = dV_\phi - dV_\phi^e \,;\, dV_s^{ne} = dV_s - dV_s^e \tag{18}$$

The non-elastic increment of pore volume $dV_\phi^{ne}$ can be calculated from the definition of the porosity (equation (1)) and knowing that $V_\phi = V - V_s$.

$$dV_\phi^{ne} = dV_\phi - dV_\phi^e = V\left[-d\varepsilon^{ne} + (1-\phi)d\varepsilon_s^{ne}\right] \tag{19}$$

From (19) we obtain:

$$\frac{dV_\phi^{ne}}{V_\phi} = \frac{-1}{\phi}\left[d\varepsilon^{ne} - (1-\phi)d\varepsilon_s^{ne}\right] \tag{20}$$

Using equation (20), equation (3) is re-written with the additional contribution of the non-elastic volume changes:

$$\begin{aligned}-\frac{dV}{V} &= \frac{d\sigma_d}{K_d} + \frac{dp_f}{K_s} + d\varepsilon^{ne} \\ -\frac{dV_\phi}{V_\phi} &= \frac{d\sigma_d}{K_p} + \frac{dp_f}{K_\phi} + \frac{d\varepsilon^{ne}}{\phi} - \frac{1-\phi}{\phi}d\varepsilon_s^{ne}\end{aligned} \tag{21}$$

Using equations (7) and (21) the following relation is obtained for the variations of the porosity:

$$\frac{d\phi}{\phi} = -\frac{1}{\phi}\left(\frac{1-\phi}{K_d} - \frac{1}{K_s}\right)d\sigma_d + \left(\frac{1}{K_s} - \frac{1}{K_\phi}\right)dp_f - \frac{1-\phi}{\phi}\left(d\varepsilon^{ne} - d\varepsilon_s^{ne}\right) \tag{22}$$

Replacing equation (21) and then equation (6) in equation (13) the following relation is found for the case of $d\sigma = 0$:

$$\rho_f \frac{dp_f}{dt}\left[\phi\left(\frac{1}{K_f} - \frac{1}{K_\phi}\right) + \left(\frac{1}{K_d} - \frac{1}{K_s}\right)\right] - \rho_f \frac{d\varepsilon^{ne} - (1-\phi)d\varepsilon_s^{ne}}{dt} + \frac{\partial q}{\partial z} = 0 \tag{23}$$

Replacing equations (15) and (17) in equation (23) the following expression is found:

$$\frac{dp_f}{dt} - \beta_u \frac{d\varepsilon^{ne} - (1-\phi)d\varepsilon_s^{ne}}{dt} = \frac{\beta_u}{\rho_f} \frac{\partial}{\partial z}\left(k \frac{\rho_f}{\mu_f} \frac{\partial p_f}{\partial z}\right) \tag{24}$$

If the non-elastic strains of the solid phase are neglected with respect to the total non-elastic strains, which is a common assumption for most geomaterials, equation (24) can be simplified in the following form:

$$\frac{dp_f}{dt} - \beta_u \frac{d\varepsilon^{ne}}{dt} = \frac{\beta_u}{\rho_f} \frac{\partial}{\partial z}\left(k \frac{\rho_f}{\mu_f} \frac{\partial p_f}{\partial z}\right) \tag{25}$$

## 3  Experimental program

A class G oil well cement is used to prepare the cement paste with a water to cement ratio $w/c = 0.44$. The fresh paste is conserved in 14cm cubic moulds for four days in lime saturated water at 90°C temperature. This temperature is chosen in order to reproduce the curing conditions of a cement lining installed in a deep oil well. After four days the temperature is gradually reduced in order to prevent any cracking of the blocs due to a sudden temperature change. Then, the blocs are cored and cut to obtain cylindrical samples with 38 mm diameter and 76 mm length. The two ends of the cylindrical samples are rectified to obtain planar surfaces perpendicular to the axis. After the sample preparation phase, the samples are cured for at least three months in a bath containing a chemically neutral solution towards the cement paste under a controlled temperature of 90°C. Before performing a test, the temperature of the sample is reduced slowly to prevent any thermal cracking. The porosity of the studied cement paste was evaluated by mercury intrusion porosimetry to be equal to $\phi = 0.26$. More details about the preparation of the samples can be found in Ghabezloo *et al.* [12].

The triaxial cell used in this study can sustain a confining pressure up to 60 MPa. It contains a system of hydraulic self-compensated piston. The loading piston is then equilibrated during the confining pressure build up and directly applies the deviatoric stress. The axial and radial strains are measured directly on the sample inside the cell with two axial transducers and four radial ones of LVDT type. The confining pressure is applied by a servo controlled high pressure generator. Hydraulic oil is used as confining fluid. The pore pressure is applied by another servo-controlled pressure generator with possible control of pore volume or pore pressure. More details on this triaxial equipment are found in Sulem and Ouffroukh [20] and in Ghabezloo and Sulem [21].

After the curing period the samples which are stored in a fluid at 90°C can be considered as completely saturated. The process of installing the sample inside triaxial cell may cause a partial de-saturation of the sample. For this reason a seven days re-saturation phase is performed inside the triaxial cell. During this period, the sample is maintained under a confining pressure equal to 1.2MPa and a back fluid pressure equal to 1.0MPa is applied to the sample while the volume of the fluid injected in the sample is monitored.

As explained earlier, the present permeability test is performed after an undrained hydrostatic compression test and uses the induced excess pore pressure inside the sample for the evaluation of the permeability. One can see several advantages in this method. First it gives access to the permeability of the sample under different states of stress (including high levels of stress and pore pressure) with a single test and gives thus an estimate of the porosity permeability relationship. Moreover, in creeping materials with low permeability, the standard tests cannot give accurate information on the short-term behaviour but only allows an estimation of the long-term permeability. An other marginal advantage is that one high capacity pressure generator for the application of the confining pressure and a low capacity one for the saturation process and the application of the pore pressure at one end of the sample are needed, while, in a conventional permeability test two additional high capacity pressure generators are needed for application and control of the pore pressure at the ends of the sample. The undrained loading is performed using a loading rate equal to 0.1MPa/min. This loading rate is slow enough to ensure the homogeneity of the generated pore pressure inside the sample. After the undrained loading phase, the pore pressure at one end of the sample is reduced instantaneously to a low value (1.0MPa) which causes a fluid flow out of the sample. The other end the drainage system is closed and the pore pressure is monitored using a pore pressure transducer. The volume of the drainage and pressure measurement system at the closed side of the sample influences the measurement of the pore pressure. In our triaxial cell the total volume of the drainage system for an undrained loading is equal to 2300mm$^3$ which is a relatively small volume. 400mm$^3$ of this volume is connected to the lower side of the sample and the remaining 1900mm$^3$ is connected to the upper side. In order to minimize the effect of the dead volume of the drainage system, the fluid flow is established at the upper side which has a greater dead volume and the pressure measurement is performed at the lower side which has a smaller dead volume. Comparing this small dead volume (400mm$^3$) to the volume of the pore space of the sample (22410mm$^3$), one can see that the correction of the pore pressure measurement as derived in Ghabezloo and Sulem [21] can be neglected.

The instantaneous reduction of the pore pressure at the upper end of the sample causes a high pressure gradient and a flow of the pore fluid out of the sample. The gradual decrease of the pore pressure inside the sample, while the confining pressure is kept constant, increases the effective stress in a non-uniform manner along the height of the sample which results in non-uniform deformations. Depending on the nature of the tested material and on the magnitude of effective stress change, these deformations may include a non-elastic part. Non-elastic strains have a significant effect on the response of the sample and should be taken into account in the analysis of the test results. During the test, the radial deformations are measured using four LVDT transducers in the middle of the sample. These measured deformations along with the pore pressure measurements are used in the back analysis of the permeability as shown in the next section.

Three permeability tests (A, B and C) were performed. The test results are presented on Figure (1) where the pore pressure measured at the lower end of the sample and the radial strain measured in the middle of the sample are plotted as functions of time. The three tests correspond to different confining pressures and different initial pore pressure inside the sample. Tests B and C have been performed on the same sample. At the end of the test B the sample is unloaded in drained condition and then reloaded in undrained condition up to 50MPa, then test C is performed. One can see in Figure (1) that for all the tests the rate of pore pressure decrease reduces significantly after about 7 hours. One can also observe that despite the small change of effective stress after 7 hours, the measured strains continue to increase significantly. This phenomenon may be attributed to the creep of the hardened cement paste under the applied effective stress.

## 4 Analysis of the results

The results obtained in the performed test are back analysed to evaluate the permeability of the sample and its changes during the test. The analysis is performed using equation (25) which takes into account the effect of non-elastic strains produced during the test. As one can see in this equation, the poroelastic properties of the tested material must be known to evaluate the $\beta_u$ parameter (equation (17)). The results of a previous experimental study on the poromechanical behaviour of the considered hardened cement paste are presented in Ghabezloo *et al*. [12]. The unjacketed modulus $K_s$ was evaluated equal to 21GPa. Several drained hydrostatic compression tests revealed the stress dependency of the drained bulk modulus $K_d$ of the hardened cement paste. The test results showed the degradation of the elastic bulk modulus with the effective stress increase:

$$K_d = 8.69 - 0.087\sigma_d \qquad \left(K_d : \text{GPa}, \sigma_d : \text{MPa}\right) \tag{26}$$

The microscopic observation of the samples after the drained tests showed that this damage phenomenon is caused by the micro-cracking of the hardened cement paste, even under hydrostatic loading, which can be attributed to the heterogeneity of the microstructure of this material. The parameter $K_\phi$, as explained above, is very difficult to evaluate and is commonly taken equal to the unjacketed modulus $K_s$. It can be shown that the deviation of this parameter from the unjacketed modulus $K_s$ does not have a great influence on the poromechanical formulations (Zimmerman *et al.*, [33]). The physical parameters of pore fluid, $\rho_f$, $\mu_f$ and $K_f$ are taken equal to the ones of pure water and their variations with pore pressure are taken into account (Spang [35]).

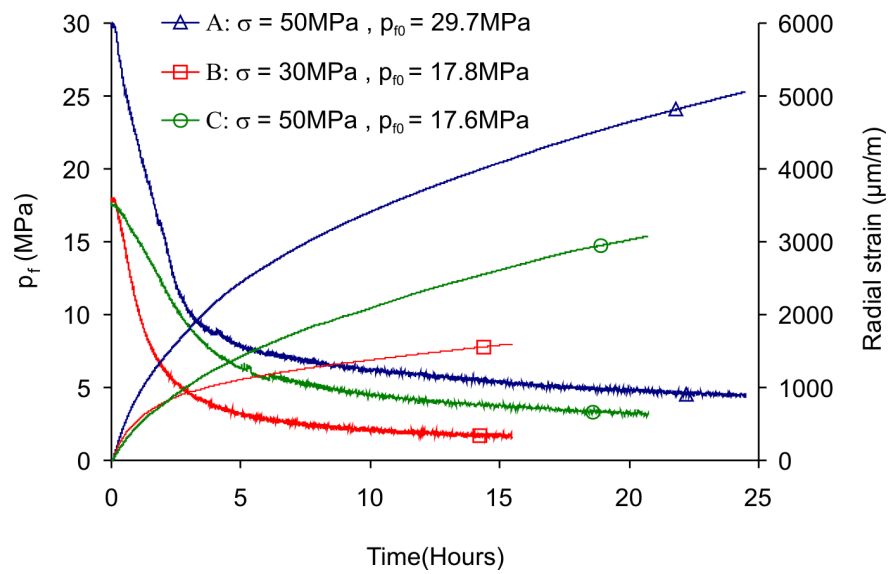

**Figure 1- Results of the permeability tests: Evolution of the pore pressure measured at the lower end of the sample and of the radial strain measured at the middle of the sample during the test.**

The elastic strains due to the variations of the pore pressure can be calculated easily using equation (3). The LVDTs measurements give the radial deformations in the middle of the sample. The measured radial strains are decomposed into an elastic and a non-elastic part. Due to the non-uniformity of the pore pressure, different points of the sample undergo different deformation. In order to estimate these strains in all points of the sample a constitutive law must be assumed for the non-elastic part of the strains. The parameters of this model are calibrated from the strains measured in the middle of the sample. Here we assume that the non-elastic part of the strains is of viscoelastic type and we choose a simple model, composed of a system of parallel spring and dashpot. We also assume that the creep rate is controlled by Terzaghi effective stress (i.e. differential pressure $\sigma_d$).

$$\frac{d\varepsilon^{ne}}{dt} = a\left(b\sigma_d - \varepsilon^{ne}\right) \tag{27}$$

In equation (27), $a$ and $b$ are two model parameters; $(ab)^{-1}$ is the viscosity and $b$ is the compressibility of the underlying dashpot and spring model. It should be noted at this point that, as the tested samples have experienced different loading paths before the permeability tests, their creep behaviour may be different so that $a$ and $b$ should be calibrated for each test.

The variations of the stress state and also the deformations of the sample during the test can modify its permeability. In order to evaluate and take into account the variation of the permeability during the test, we can assume that the permeability varies either with the stress state or with the porosity changes. Assuming that the permeability is controlled by the stress state is appropriate if the deformation process is elastic. In that case, an appropriate effective stress law for the variation of the permeability has to be established, as explained in Ghabezloo et al. [36]. On the other hand, for a material which exhibits time-dependent deformation, such an assumption is not appropriate as for example, the permeability changes during a creep test due to the deformation process whereas the stress state remains constant. Thus, for complex deformation process, empirical permeability-porosity laws are commonly used. We choose here a power law $k/k_0 = (\phi/\phi_0)^\alpha$, so that we can write:

$$\frac{dk}{k} = \alpha \frac{d\phi}{\phi} \tag{28}$$

As mentioned before, the coefficient $\alpha$ may be constant or variable, integer or non-integer, according to the properties of the material and of the evolution process. For the sake of simplicity, in the back analysis of the results a constant integer coefficient is assumed. The exponent $\alpha$ and the initial permeability $k_0$ are evaluated in the back analysis of the experimental results. Before the beginning of the permeability evaluation test, the permeability of the sample is first modified by the applied undrained loading. The undrained test performed on this sample with three unloading-reloading cycles at different levels of confining pressure, as presented in Ghabezloo et al. [12], does not show any irreversible strains in unloading. Consequently the variations of the sample porosity caused by the applied undrained loading can be easily evaluated using equation (8) by knowing the final values of confining pressure and pore pressure. The initial porosity is equal to $\phi_0 = 0.26$ and the drained bulk modulus $K_d$ is calculated using equation (26) for the average effective stress during the undrained loading. Using these parameters, the porosity change and the resulting permeability change caused by the initial undrained loading can be evaluated (equation (28)).

The back analysis of the test results using equation (25) is performed using a finite difference numerical scheme. The first analysis is performed on test A. The parameters $a$ and $b$, the exponent $\alpha$ and the initial permeability $k_0$ are evaluated using the least square method. The error between the computed results and the measured data is calculated as the sum of errors corresponding to the pore pressure and the radial strain. For each iteration, the parameters are changed manually and the sum of square of the difference between the measured and the calculated values is calculated on a set of points. The best fitting parameters are then evaluated by minimizing the calculated error. For the test A the following parameters are found: $a = 2 \times 10^{-5} \text{ s}^{-1}$, $b = 3.2 \times 10^{-4} \text{ MPa}^{-1}$, $\alpha = 11$ and $k_0 = 1.07 \times 10^{-19} \text{ m}^2$. The results of the numerical simulation together with the experimental data are presented in Figure (2). One can see that the measured radial strains are well reproduced in the simulation using the assumed viscoelastic model and the calibrated parameters. The measured pore pressure at the lower end of the sample is also well reproduced by the model.

The variation of calculated permeability at the centre of the sample during the test is presented in Figure (3). One can see an initial permeability reduction from $1.07 \times 10^{-19} \text{ m}^2$ to $1.03 \times 10^{-19} \text{ m}^2$ under the effect of the initial undrained loading followed by a progressive reduction to $6.34 \times 10^{-20} \text{ m}^2$ at the end of the test. We observe that the undrained loading phase does not modify much the permeability of the sample. By doing this analysis we obtain a permeability porosity relationship for the tested sample using the results of a single transient test under the following form:

$$k = 1.07 \times 10^{-19} \left( \frac{\phi}{0.26} \right)^{11} \tag{29}$$

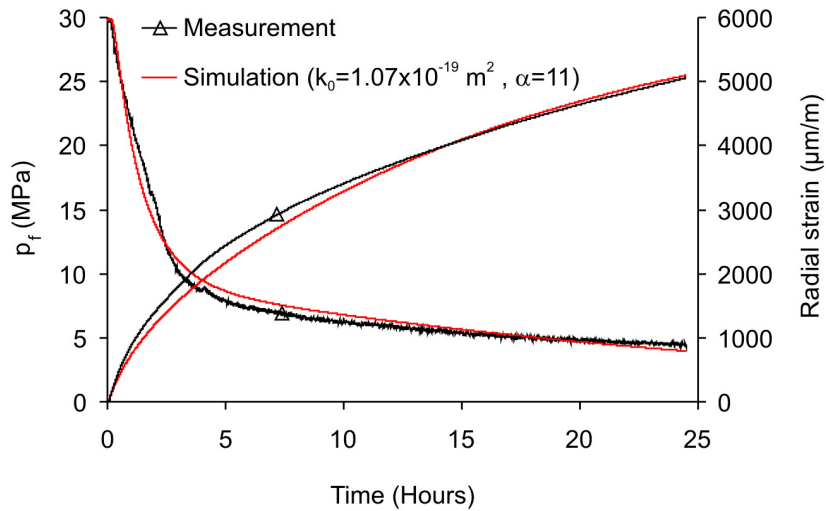

**Figure 2- Test A: Test results and numerical simulation.**

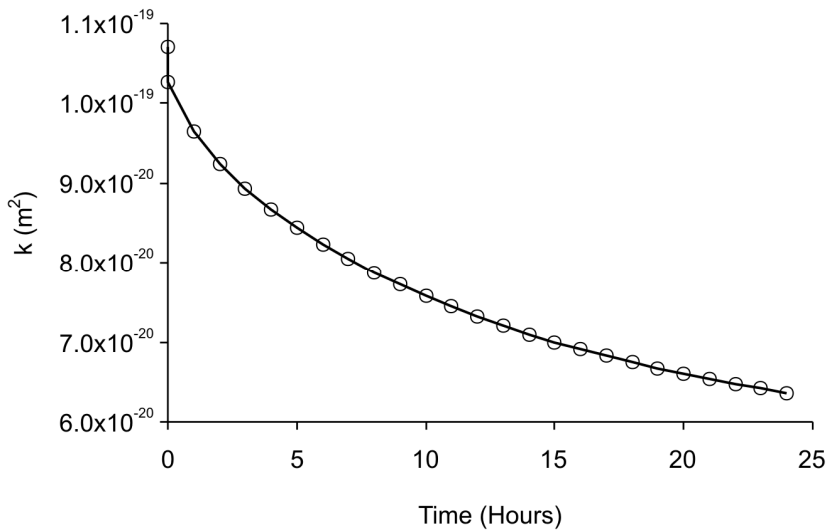

**Figure 3- Test A: Evolution of the calculated permeability at the centre of the sample.**

The above computation takes into account both the effect of non-elastic deformation and the effect of permeability change with porosity. In order to investigate separately these two effects, two additional simulations are performed. In the first one, the analysis is performed by neglecting the effect of non-elastic strains during the test and by considering a constant permeability (case 1). In the second one, the effect of non-elastic strains is taken into account, keeping a constant permeability (case 2). For each case the required parameters are evaluated using the least square algorithm. The results are presented on the Figure (4) and compared with the results of the complete model (case 3). For case 1, significant differences between the experimental results and the computed ones are observed. Especially the important difference between the measured and the calculated radial strain shows that it is essential to take into account the effect of non-elastic strains in the analysis of the test results. The results of case 2 show that an acceptable response can be obtained by considering the effect of non-elastic strains even with a constant permeability. Case 3 is identical to the analysis presented on Figure (2) and results in an even better compatibility with the experimental results in particular for the permeability curve in the last part of the test. For case 2, a constant permeability is evaluated equal to $8.30 \times 10^{-20} \, m^2$ and can be seen as an average permeability of the deforming sample during the test. The relatively small difference between the results of case 2 and case 3 is due to the small reduction of the porosity and consequently of the permeability of the sample during the test (Figure (3)).

Two main physical mechanisms occur in the course of the loading process, micro-cracking as mentioned above and compaction. The reduction of the permeability during the test shows that the compaction process is dominant and that the net result of these two mechanisms is a reduction of the permeability. Moreover, one can infer that the induced micro-cracks remain closed under stress and do not influence the permeability significantly.

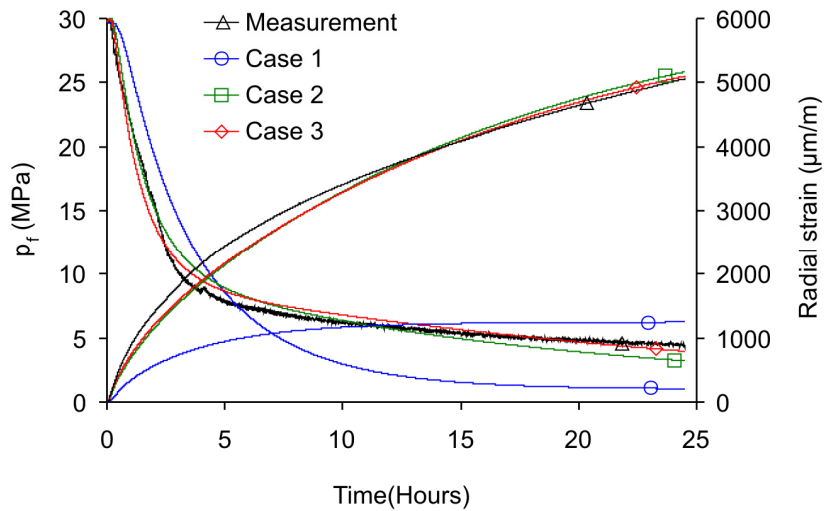

**Figure 4- Test A: Effect of non-elastic strains and porosity dependent permeability: Case 1: Elastic solution, constant permeability; Case 2: Non-elastic solution, constant permeability; Case 3: Non-elastic solution, porosity dependent permeability.**

For the analysis of the results of the tests B and C we assume that the coefficient $\alpha$ is only a function of the compaction process and consequently is the same for all tests, equal to 11. For each of the tests B and C, the coefficients $a$ and $b$ of equation (27) and the initial permeability $k_0$ are evaluated to find the best accordance between the simulation and the test results. For test B the following parameters are found: $a = 4.1 \times 10^{-5}\,\mathrm{s}^{-1}$, $b = 1.3 \times 10^{-4}\,\mathrm{MPa}^{-1}$ and $k_0 = 1.20 \times 10^{-19}\,\mathrm{m}^2$. For this sample we thus obtain the following permeability law:

$$k = 1.20 \times 10^{-19} \left(\frac{\phi}{0.26}\right)^{11} \qquad (30)$$

At the end of this test, the permeability of the node at the centre of the sample is $1.02 \times 10^{-19}\,\mathrm{m}^2$. As mentioned above, this sample is then unloaded in drained condition and test C is performed on the same specimen. For test C the evaluated parameters are: $a = 2.4 \times 10^{-5}\,\mathrm{s}^{-1}$, $b = 1.8 \times 10^{-4}\,\mathrm{MPa}^{-1}$ and $k_0 = 9.40 \times 10^{-20}\,\mathrm{m}^2$. As expected, the initial permeability of the sample calculated for test C is very close to the final permeability of the sample in test B which gives confidence in the approach. The results are presented in Figures (5) and (6) where a good compatibility between the simulation and the test results can be observed. In Figure (6) a small pore pressure increase can be observed in the beginning of the test C in the results of the numerical simulation. Due to the high value of the initial effective stress in this test (32.4MPa) and also the low permeability of the sample, in the beginning of the test the effect of the creep deformations is more important than the effect of the pore pressure diffusion. This explains the small initial pore pressure increase obtained in the simulation.

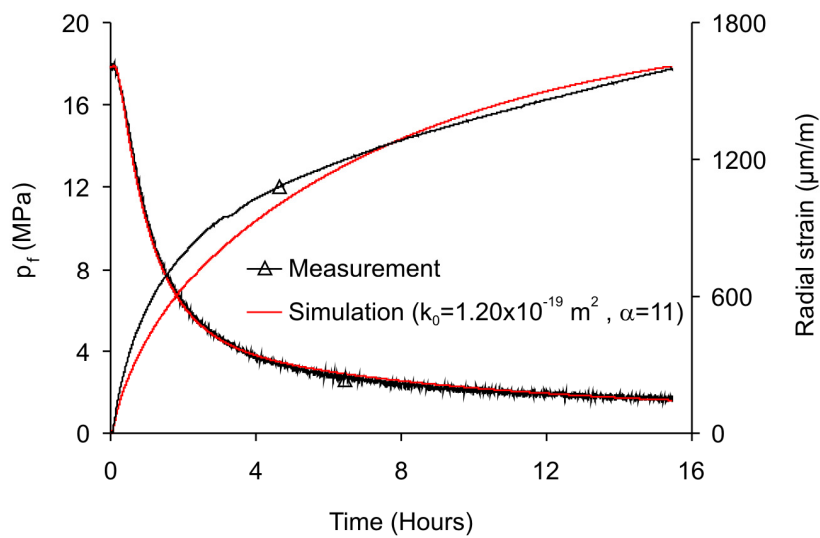

**Figure 5- Test B: Test results and numerical simulation.**

The evaluated value of the porosity sensitivity exponent $\alpha$, equal to 11, is a relatively high value comparing with the range of variations of $\alpha$ for other geomaterials. Such a high value of $\alpha$ is usually attributed to a pore structure containing relatively large, pressure-insensitive nodal pores connected to each other by narrow pressure-sensitive throats (Bernabé *et al.* [14]) or to a pore structure in which pocket-like pores contribute much to the porosity and little to the permeability (David *et al.* [17]). This latter description seems to be mostly appropriate for the case of a permeability-porosity relationship which is established for different samples of a porous material with different porosities. Based on the idea of Bernabé *et al.* [14], a high value of $\alpha$ means that as a result of the compaction process, the reduction of the effective porosity is more important than the reduction of the non-effective porosity, or that some part of the effective porosity is transformed into non-effective porosity. We have shown that for two samples of a given material, the permeability evolves with the porosity with the same value of $\alpha$, mostly independently of the level of effective stress or creep deformations in the three performed tests. This shows that in the absence of a difference between the pore shape and geometry of the samples, the permeability variation is controlled by the process of porosity evolution, which is the same in the different tests. This result is in accordance with the idea of Bernabé *et al.* [14], as explained here above.

The obtained permeability-porosity relationship with an exponent $\alpha$ equal to 11 should not be seen as a general permeability-porosity law for cement paste but specific to the cement used in this study and to the specific hydrostatic compaction process. This relation could in principle be combined with other empirical relationships found in the literature which describe the evolution of the permeability of the cement paste with other parameters such as $w/c$ (water to cement ratio), the pore size distribution etc…
For example Breysse and Gérard [37] performed a statistical analysis on the values of permeability of different cement pastes prepared with different $w/c$ ratios as reported in the literature and found a power law for the relationship between the permeability and $w/c$. Higher values of $w/c$ result in a cement paste with a higher initial porosity and thus a higher initial permeability. Breysse and Gerard relationship can be used to evaluate the initial permeability to be put in our permeability evolution law. On the other hand empirical relationships can be found in the literature to estimate the initial porosity as a function of $w/c$ (Taylor [38]).

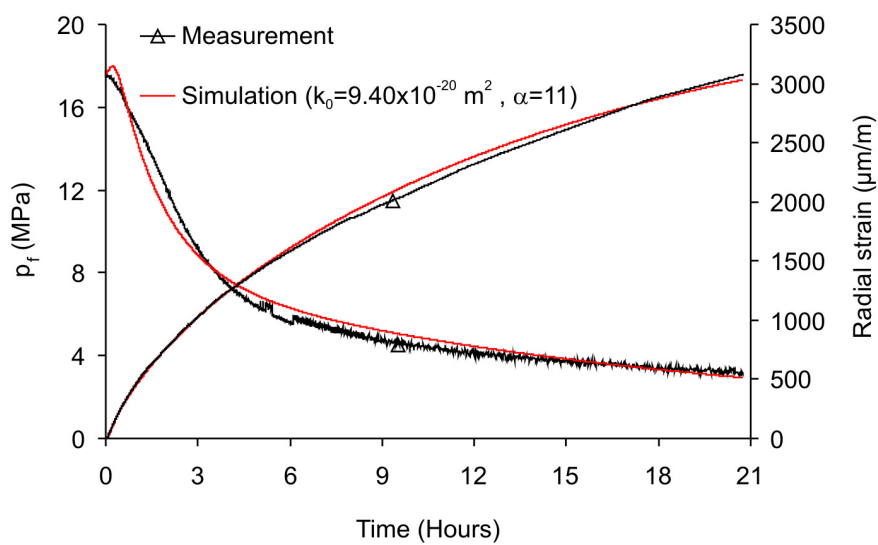

**Figure 6- Test C: Test results and numerical simulation.**

# 5 Conclusions

The permeability-porosity relationship for a hardened class G oil well cement paste is evaluated experimentally using a single transient test. The excess pore pressure generated in the sample after an undrained hydrostatic compression test is used for the evaluation of the permeability. This pressure is released at one end of the sample and the variations of the pore pressure at the other end and also the radial deformations in the middle of the sample are measured during the dissipation of the pore pressure. The major advantage of the proposed method as compared to the classical steady state or transient permeability measurement methods, which are performed at a given level of stress, is that one can directly access to the permeability-porosity relationship in a single test. The test results are back analysed to evaluate a power-law permeability-porosity relationship. The creep of the sample during the test and its effect on the measured pore pressure and deformations are introduced in the formulation and are taken into account in the back analysis of the results. For a non creeping material, the same type of analysis can be used to assess a permeability-effective stress relationship. The exponent of the power-law is evaluated equal to 11 in one of the tests and is shown to be the same for the other tests with different initial stress and pore pressure levels. Consequently we can conclude that this coefficient is mostly dependent upon the compaction process which modifies the porosity of the material in accordance with the idea presented by Bernabé *et al*. [14].

It should be emphasized that the obtained power-law relationship describes the variations of the permeability of a given hardened cement paste with a given initial porosity submitted to a hydrostatic loading. In order to extrapolate this law to other cement pastes, prepared with different *w/c* ratio, one can use empirical relationships existing in the literature which relate the initial permeability and the initial porosity to this parameter.

# 6 Acknowledgments

The authors gratefully acknowledge TOTAL for supporting this research. They wish also to thank Yves Bernabé and Ahmad Pouya for fruitful comments and discussions and François Martineau for his assistance in the experimental work.